%% file: main.tex
\documentclass{article}

\usepackage{PRIMEarxiv}
\usepackage{algpseudocode}
\usepackage{algorithm,algorithmicx}
\usepackage{caption}
\usepackage{adjustbox}
\usepackage{amsmath}
\usepackage{xcolor}
\usepackage[utf8]{inputenc} 
\usepackage[T1]{fontenc}    
\usepackage{biblatex} 
\usepackage{float}
\usepackage{hyperref}       
\usepackage{url}            
\usepackage{booktabs}       
\usepackage{amsfonts}       
\usepackage{nicefrac}       
\usepackage{microtype}      
\usepackage{lipsum}
\usepackage{fancyhdr}       
\usepackage{graphicx}       
\graphicspath{{media/}}     
  
\title{TP-Aware Dequantization}

\author{
   Adnan Hoque \\
  IBM T.J. Watson Research Center \\
  Yorktown Heights, NY, United States\\
  \texttt{adnan.hoque1@ibm.com} \\
  \And
  Mudhakar Srivatsa \\
  IBM T.J. Watson Research Center \\
  Yorktown Heights, NY, United States\\
  \texttt{msrivats@us.ibm.com} \\
   \And
  Chih-Chieh Yang \\
  IBM T.J. Watson Research Center \\
  Yorktown Heights, NY, United States\\
  \texttt{chih.chieh.yang@ibm.com} \\
   \And
  Raghu Ganti \\
  IBM T.J. Watson Research Center \\
  Yorktown Heights, NY, United States\\
  \texttt{rganti@us.ibm.com} \\
}

\begin{document}
\maketitle

\begin{abstract}
In this paper, we present a novel method that reduces model inference latency during distributed deployment of Large Language Models (LLMs). Our contribution is an optimized inference deployment scheme that address the current limitations of state-of-the-art quantization kernels when used in conjunction with Tensor Parallel (TP). Our method preserves data locality in GPU memory access patterns and exploits \textit{a priori} knowledge of TP to reduce global communication. We demonstrate an up to 1.81x speedup over existing methods for Llama-70B and up to 1.78x speedup for IBM WatsonX's Granite-20B MLP layer problem sizes on A100 and H100 NVIDIA DGX Systems for a variety of TP settings.
\end{abstract}

\keywords{Deep Learning  \and Foundation Models \and Quantization \and Tensor Parallel \and GPU \and Optimization \and LLM \and Transformers \and Distributed Systems}

\section{Introduction}

Given the recent advancement of LLMs, deployment optimizations are becoming more crucial as the size of state-of-the-art LLMs increase in scale. As these these models continue to grow, so does the need to optimize the increasingly parallel and increasingly distributed workload requirements of modern-day deep learning inference. Strategies like GPTQ \cite{frantar_gptq_2023} and Tensor Parallel (TP) \cite{noauthor_tensor_nodate} are hence essential in achieving high-throughput performance. Our method is motivated by several key properties of GPTQ, TP and General Matrix Multiplication (GEMM). We build on these existing methods and present a key innovation that helps maximize memory throughput and reduce latency. Our method shows up to a 1.81x speedup on Llama-70B and up to a 1.78x speedup on Granite-20B MLP layer problem sizes. We achieve this by reducing global communication and enforcing data locality.

\subsection{Motivation}

We begin by introducing the grouping scheme in GPTQ-style quantization. The \textit{group size} defines the quanta that the quantization strategy will be performed on. Thus, every \textit{group size} number of input channels in the \textit{K x N} weight matrix will share the same quantization metadata (scales and zeros). This introduces the need to spatially relate the weight matrix to the aforementioned metadata. This mapping must be remembered, as it will be used during deployment when performing \textit{dequantization}. If we were to proceed with the basic formulation of GPTQ, the rows of the weight matrix are related to the metadata by a group index array. In this paper, we will refer to this basic formulation of the group index array as the \textit{naive} approach and it can be evaluated as the following, where we let \textit{G} equal the group size:

\begin{equation} \label{eq:1}
g_{\text{idx\_naive}}[i] = \left\lfloor \frac{i}{\text{G}} \right\rfloor, \quad \text{for}\  i = 0, 1, 2, \ldots, K-1
\end{equation}

Notably, this approach does not consider further optimizations introduced by the GPTQ authors. Our method targets the complexities that arrive when considering the "Activation Order" optimization. This parameter was introduced by the authors to improve model accuracy \cite{lbonne_ml_2023} and is triggered through an optional flag act\_order in the GPTQ package. When this flag is set to True, the quantization process will behave differently. This optimization reduces overall quantization error by processing the rows of the weight matrix in a way that respects the significance of the weights impacts on model accuracy. Thus, the weights with higher impact on accuracy will incur less quantization error, and vice versa for weights with lower impact on accuracy. The implication is then, with act\_order (also known as desc\_act in different packages), the rows of the weight matrix are reordered based on the properties discussed above and notably, the new ordering \textit{must} be respected when performing inference during dequantization for each row of the weight matrix and it's corresponding metadata. This transformation is realized through the group index array. Here, we use a random permutation function $\phi$ to emulate an arbitrary reordering. The new group index array can then be expressed by the following expression:

\begin{equation} \label{eq:2}
\text{Let}\ \phi: \{0, 1, \ldots, K-1\} \rightarrow \{0, 1, \ldots, K-1\} \text{\ be \ a \ random \ permutation \ function.}
\end{equation}

\begin{equation} \label{eq:3}
g_{\text{idx\_actorder}}[i] = \left\lfloor \frac{\phi(i)}{\text{G}} \right\rfloor, \quad \text{for}\ i = 0, 1, 2, \ldots, K-1
\end{equation}

This approach however, introduces performance overhead due to the sub-optimal memory access pattern of having to frequently reload quantization metadata \cite{lbonne_ml_2023} due to the \textit{unordered} nature of the group index array. Further, when this optimization is used in a TP setting this method incurs communication overhead, a detail we expand on in the next section. Our method is then, a way to reap the benefits of the accuracy boost provided by the act\_order optimization without reducing memory throughput and incurring additional communication overhead.

\section{Method}

\subsection{Enforcing Data Locality}

\begin{figure}[H]
    \centering
    \begin{minipage}[b]{0.48\textwidth}
        \includegraphics[width=\textwidth]{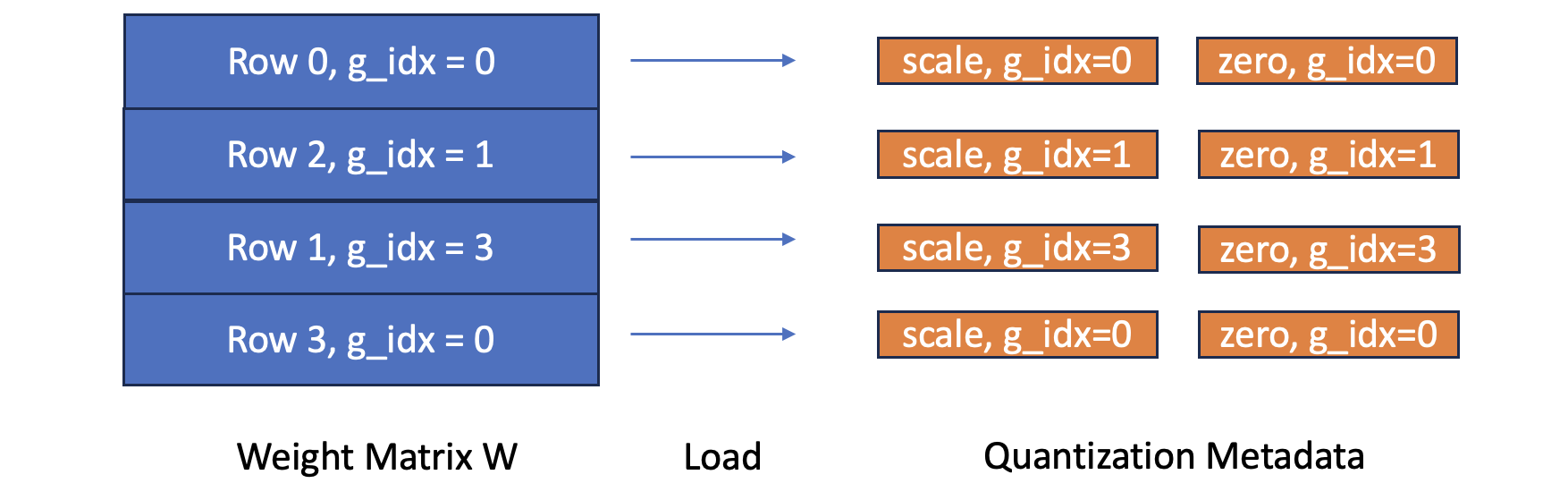}
        \caption{Naive Load with Activation Order Flag}
        \label{fig:unordered}
    \end{minipage}
    \hfill
    \begin{minipage}[b]{0.48\textwidth}
        \includegraphics[width=\textwidth]{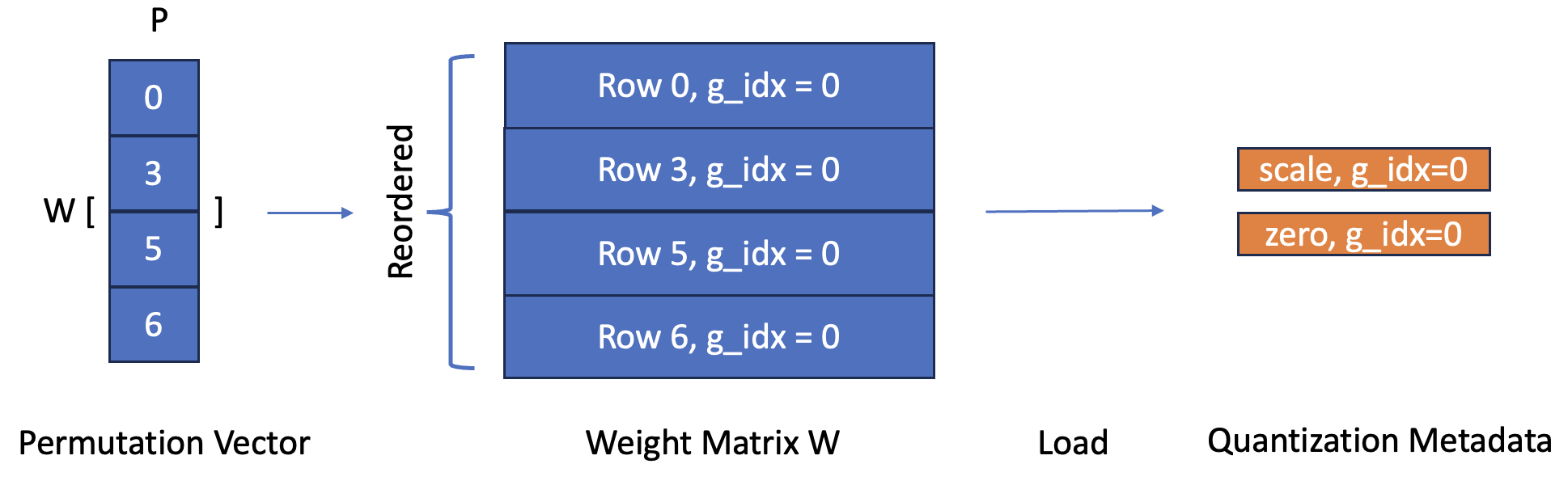}
        \caption{Optimized Load with Activation Order Flag}
        \label{fig:ordered}
    \end{minipage}
\end{figure}

Popular software packages that implement the GPTQ algorithm \cite{william_panqiweiautogptq_2023} will store the weights of the model on disk without including knowledge of the ordering suggested by Equation \ref{eq:3}. Thus, we have a choice to use the mapping in Equation \ref{eq:3} to correctly map rows of the weight matrix to it's corresponding metadata as seen in Figure \ref{fig:unordered} during model deployment. As discussed previously, this leads to a sub-optimal memory access pattern. The method described in this section, is used in the state-of-the-art ExllamaV2 \cite{turboderp_turboderpexllamav2_2023} kernel to enforce data locality by transforming the unordered group index array resulting from the act\_order flag. To illustrate this method, we define a routine that takes as input the group index array, and returns an \textit{ordered} group index array, $g_{idx\_optimized}$ as well as it's associated permutation array \textit{P}, using the torch.argsort \cite{noauthor_torchargsort_nodate} function.

\input{equations/reorder_function}

The optimized group index array has the desirable property of enforcing data locality during model inference. Consider Figure \ref{fig:unordered} and Figure \ref{fig:ordered}. In Figure \ref{fig:ordered} we see that for multiple rows of the weight matrix \textit{W} we are able to reuse the associated metadata, as the indices have been sorted such that it \textit{guarantees} all members of the the same group are consecutive in \textit{W}.

The optimized loading scheme increases data locality and memory throughput. The reordering of the weight matrices can be performed offline, and thus the permutation array \textit{P} along with the optimized group index array are available to the model inference pipeline as global variables. If we will be deploying with TP model parallelism these parameters must be passed and sharded to our input activations \textit{X}, model weights \textit{W} and quantization metadata as they will be used to perform the correct mapping during dequantization. However, with no further optimization, this strategy that we will refer to as the \textit{Naive Algorithm}, will incur an extra AllGather communication necessitated by the observation that the output of the Column-TP layer in Figure \ref{fig:tp} will require reordering, as seen in Line ~\ref{allgather} of Algorithm \ref{alg:naive_method_spmd}, to produce a correctly aligned tensor for the subsequent GEMM operation. 

This leads us to our main contribution, which is a method to reduce global communication and thus reduce latency during model inference.

\subsection{Minimizing Global Communication}

\begin{figure}[H]
    \centering
    \begin{minipage}[b]{0.48\textwidth}
        \includegraphics[width=\textwidth]{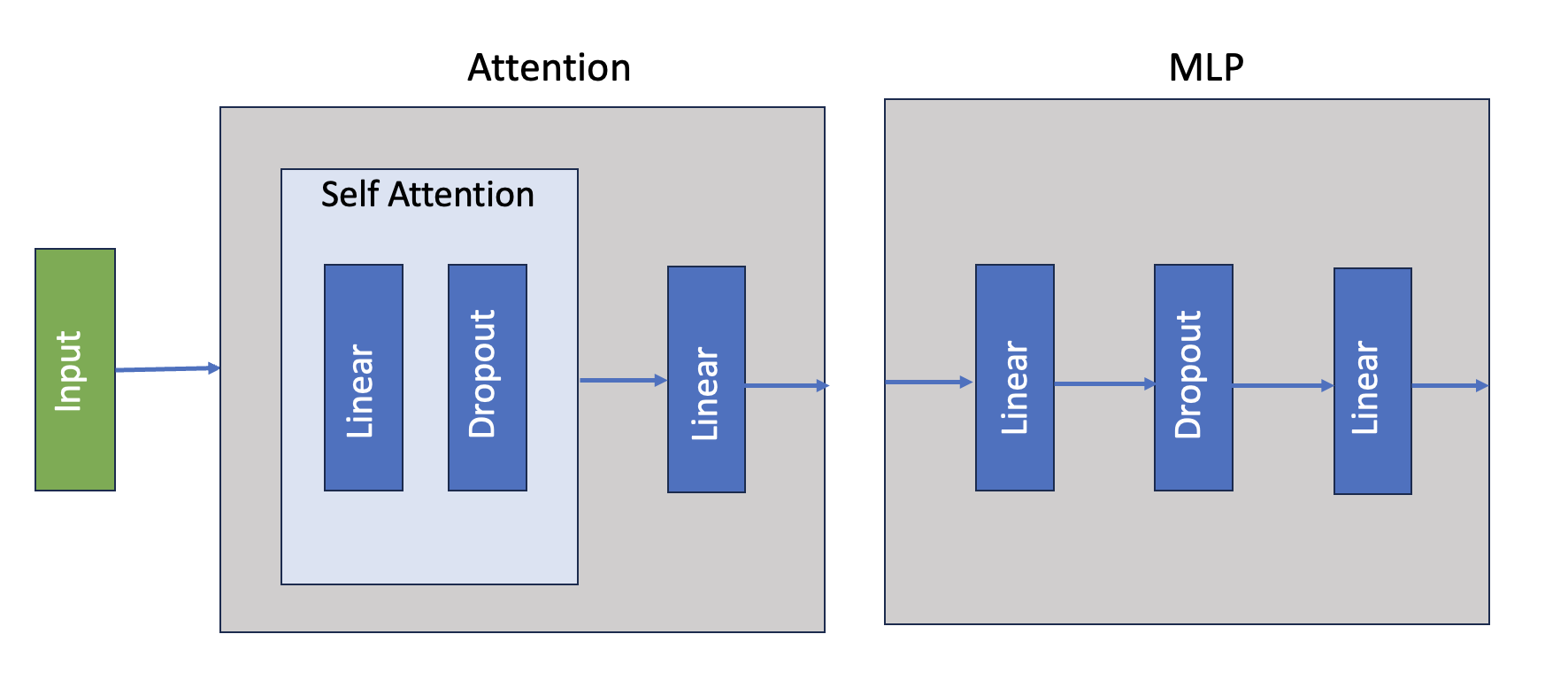}
        \caption{Transformer Block}
        \label{fig:transformer}
    \end{minipage}
    \hfill
    \begin{minipage}[b]{0.48\textwidth}
        \includegraphics[width=\textwidth]{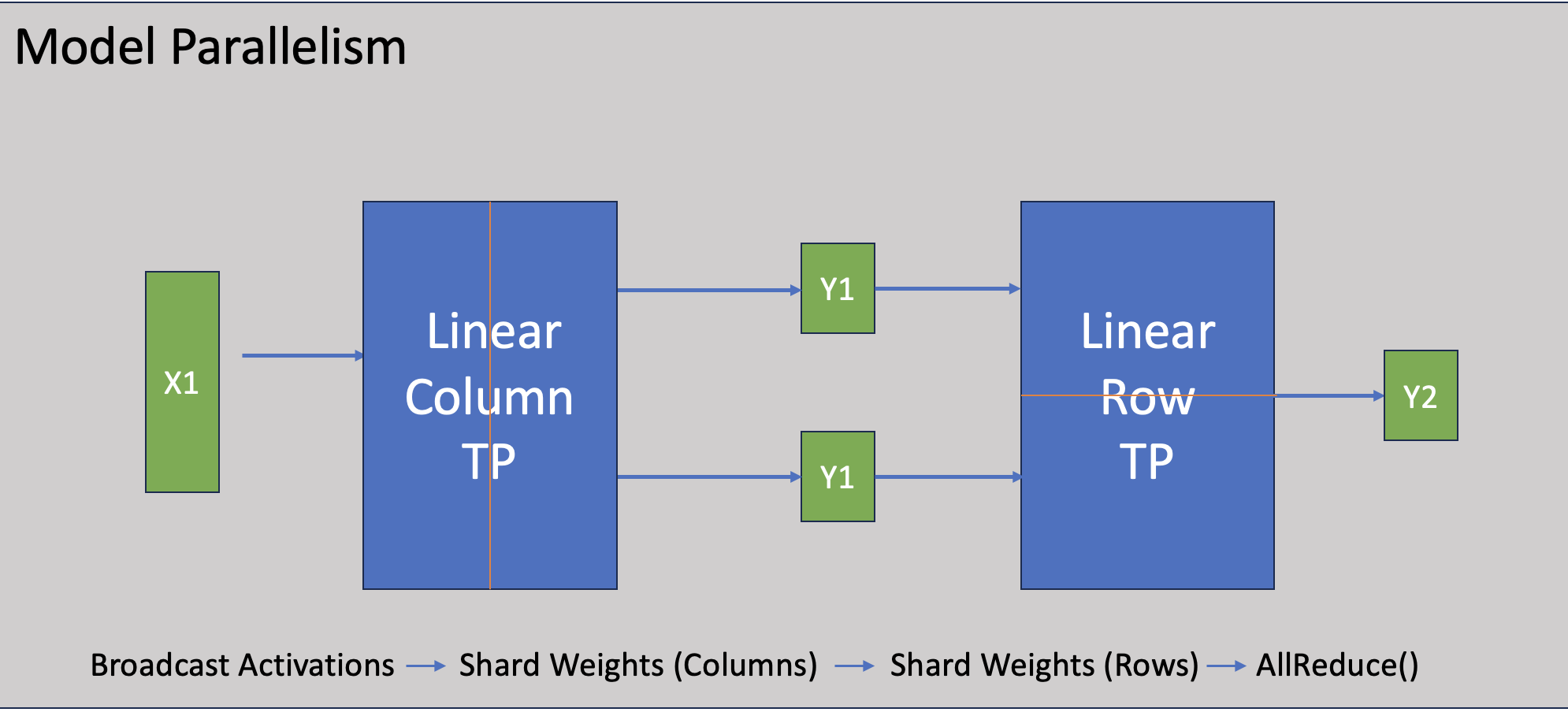}
        \caption{TP-Aware Model Parallelism}
        \label{fig:tp}
    \end{minipage}
\end{figure}

In Transformer based architectures, we typically observe that popular implementations consist of 2 linear layers in the Attention block and 2 linear layers in the Multilayer Perceptron Layer (MLP) block as depicted in Figure \ref{fig:transformer}. We follow the same model parallel strategy outlined in Megatron-LM \cite{narayanan_efficient_2021}. This is to say, our method assumes the interleaved pattern of Column-TP with Row-TP as seen in Figure \ref{fig:tp} for consecutive linear layers. In dequantization, we now have two choices. If we proceed with the optimized formulation of GPTQ with act\_order=True, we must pass either the group index array outlined in Equation \ref{eq:3} or the more optimal cache-friendly group index array from Algorithm \ref{alg:reorder}, used by the ExllamaV2 kernel.

If we proceed with the latter, during model inference, an expensive AllGather communication is required before the output shards of the first linear layer \textit{Y1} are passed to the sharded weights of the second linear layer. Our insight, is then a method that avoids this communication using the intrinsic properties of GEMM.

To motivate our method, we examine the current approach with the optimized group index array used by ExllamaV2. We first note, the interleaved Column-TP and Row-TP depicted in Figure \ref{fig:tp} will produce two sharded weight matrices, $W1$, which is split column-wise across $N$ ranks, and $W2$ which is split row-wise across $N$ ranks. During dequantization, we will produce two permutation arrays, $P1$ and $P2$. Recall these arrays are optimized in that they enforce data locality when they are used to reorder the weights and load quantization metadata (scales and zeros). Notably, these two permutation arrays are available globally and can be computed offline.

We define Algorithm \ref{alg:naive_method_spmd}, and call it the \textit{Naive Algorithm} to illustrate the above approach. We assume, activations, $X1$, the weight matrices $W1$ and $W2$ and permutation arrays $P1$ and $P2$ are available as input to the model. For a given matrix $M$, we use the following notation $M[P1, P2]$ to denote row and column permutations.

\input{equations/naive_method}

Noting the global communication required in Line~\ref{allgather} of Algorithm \ref{alg:naive_method_spmd}, we present a reordering strategy that avoids this communication across ranks and hence yields impressive latency improvements. We achieve this by by re-ordering the columns of $W1$ with the permutation array $P2$. This optimization is realized from the following insight.

By permuting the columns of $W1$ with $P2$ and then proceeding with the GEMM as is done Line~\ref{insight} of Algorithm \ref{alg:optimized_matmul_spmd} we are aligning $Y1$ in such a way it no longer needs to be permuted with $P2$ in the subsequent GEMM with $W2$. Producing this alignment, is our key contribution, as it makes it possible to avoid global communication after the column-TP layer. Note that our method as it stands, only applies to the MLP layers of the Transformer block. This is due to the fact that the sharding strategy for Attention when employing MHA (Multi-Headed Attention), MQA (Multi-Query Attention) or GQA (Group-Query Attention) motivates the need for additional tricks to avoid global communication in TP, when the weights are reordered. We call our approach that applies to the MLP layers in the Transformer Block the \textit{TP-Aware Algorithm}. 

\input{equations/optimized_tp_method}

\section{Experiments}

We conducted our experiments on a variety of TP settings and batch sizes, on highly relevant Llama-70B and Granite-20B, the flagship offering in IBM WatsonX, MLP layer sizes. We tested on enterprise A100x8 with Intel(R) Xeon(R) Platinum 8358 CPU @ 2.60 and H100x8 with Intel(R) Xeon(R) Platinum 8480 CPU @ 3.00 GHz NVIDIA DGX systems. As our algorithm avoids communication in-between column-TP and row-TP layers we use FP16 to demonstrate this benefit. For our experimental results we denote M as the batch size, K1 and N1 as the input and output features respectively of the column-TP layer and N2 as the input features of the row-TP layer and have enabled TP using 1,2,4,8 GPUs in one node. As a simplification for the analysis, in the Llama-70B test case we assume single $up_{proj}$ layer followed by $down_{proj}$ which allows us to directly compare with Granite's MLP layer. Our method can be generalized to the implementation in practice where a $gate_{proj}$ layer is also present.

\section{Llama-70B}

\begin{figure}[H]
    \centering
    \begin{minipage}{0.48\textwidth}
        \centering
        \textbf{Latency Llama-70B}\par\medskip
        \includegraphics[width=\textwidth]{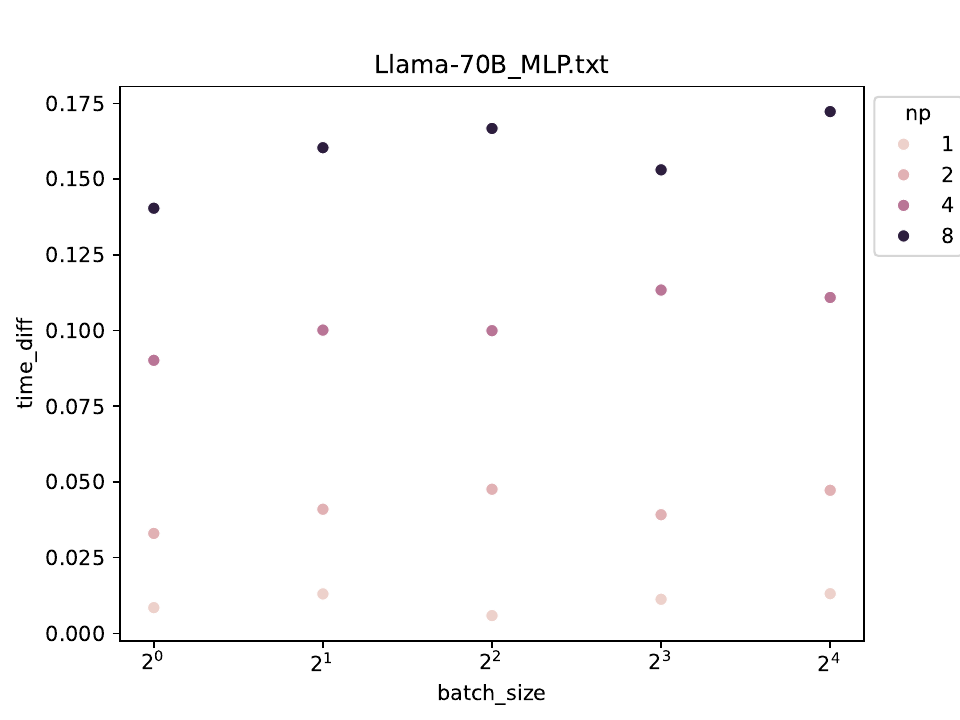}
        \caption{Latency Difference for Llama-70B, A100}
        \label{fig:first_figure}
    \end{minipage}
    \hfill
    \begin{minipage}{0.48\textwidth}
        \centering
        \textbf{Speedup Llama-70B}\par\medskip
        \includegraphics[width=\textwidth]{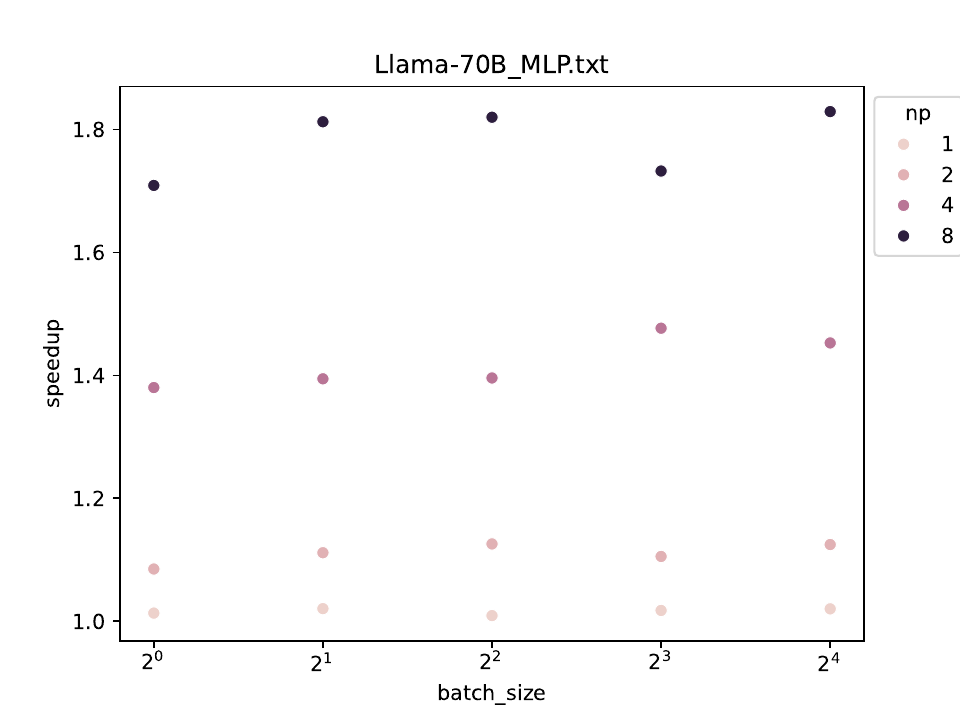}
        \caption{Speedup for Llama-70B, A100}
        \label{fig:second_figure}
    \end{minipage}
\end{figure}

\subsection{Baseline TP=1}

\begin{table}[H]
    \centering
    \begin{tabular}{|c|c|c|c|}
        \hline
        M & K1, N1, N2          & Naive Algorithm (ms) & TP Aware Algorithm (ms)   \\ \hline
        1 & (8192, 28672, 8192) & 0.696            & 0.688                 \\ \hline
        2 & (8192, 28672, 8192) & 0.694            & 0.683                 \\ \hline
        4 & (8192, 28672, 8192) & 0.685            & 0.678                 \\ \hline
        8 & (8192, 28672, 8192) & 0.706            & 0.697                 \\ \hline
        16 & (8192, 28672, 8192)& 0.710            & 0.695                 \\ \hline
    \end{tabular}
    \captionof{table}{Naive and TP Aware Algorithm Latencies for Various Batch Sizes (Llama-70B, TP=1, A100)}
    \label{tab:naive_vs_tp_various}
\end{table}

\begin{table}[H]
    \centering
    \begin{tabular}{|c|c|c|c|}
        \hline
        M & K1, N1, N2         & Naive Algorithm (ms) & TP Aware Algorithm (ms)  \\ \hline
        1 & (8192, 28672, 8192)& 0.489            & 0.481                \\ \hline
        2 & (8192, 28672, 8192)& 0.471            & 0.466                \\ \hline
        4 & (8192, 28672, 8192)& 0.474            & 0.468                \\ \hline
        8 & (8192, 28672, 8192)& 0.471            & 0.464                \\ \hline
       16 & (8192, 28672, 8192)& 0.474            & 0.468                \\ \hline
    \end{tabular}
    \caption{Naive and TP Aware Algorithm Latencies for Various Batch Sizes (Llama-70B, TP=1, H100)}
    \label{tab:naive_vs_tp_various}
\end{table}

\subsection{TP=2}

\begin{table}[H]
    \centering
    \begin{tabular}{|c|c|c|c|c|}
        \hline
        M & K1, N1, N2         & Naive Algorithm (ms) & TP Aware Algorithm (ms) & Speedup  \\ \hline
        1 & (8192, 28672, 8192)& 0.493            & 0.433               & 1.14x    \\ \hline
        2 & (8192, 28672, 8192)& 0.508            & 0.407               & 1.25x    \\ \hline
        4 & (8192, 28672, 8192)& 0.519            & 0.412               & 1.26x    \\ \hline
        8 & (8192, 28672, 8192)& 0.516            & 0.418               & 1.23x    \\ \hline
       16 & (8192, 28672, 8192)& 0.501            & 0.416               & 1.20x     \\ \hline
    \end{tabular}
    \captionof{table}{Naive and TP Aware Algorithm Latencies for Various Batch Sizes (Llama-70B, TP=2, A100)}
    \label{tab:naive_vs_tp_various}
\end{table}

\begin{table}[ht]
    \centering
    \begin{tabular}{|c|}
        \hline
        Average Speedup \\ \hline
        1.22x           \\ \hline
    \end{tabular}
    \caption{Average Speedup for Llama-70B, TP=2, A100}
    \label{tab:avg_speedup_various}
\end{table}

\begin{table}[H]
    \centering
    \begin{tabular}{|c|c|c|c|c|}
        \hline
        M & K1, N1, N2          & Naive Algorithm (ms) & TP Aware Algorithm (ms) & Speedup \\ \hline
        1 & (8192, 28672, 8192) & 0.302            & 0.283               & 1.07x    \\ \hline
        2 & (8192, 28672, 8192) & 0.316            & 0.285               & 1.11x    \\ \hline
        4 & (8192, 28672, 8192) & 0.323            & 0.286               & 1.13x    \\ \hline
        8 & (8192, 28672, 8192) & 0.320            & 0.289               & 1.11x    \\ \hline
       16 & (8192, 28672, 8192)& 0.322             & 0.289               & 1.11x    \\ \hline
    \end{tabular}
    \caption{Naive and TP Aware Algorithm Latencies for Various Batch Sizes (Llama-70B, TP=2, H100)}
    \label{tab:naive_vs_tp_various}
\end{table}

\begin{table}[H]
    \centering
    \begin{tabular}{|c|}
        \hline
        Average Speedup \\ \hline
        1.11x           \\ \hline
    \end{tabular}
    \caption{Average Speedup for Llama-70B, TP=2, H100}
    \label{tab:avg_speedup_various}
\end{table}

\subsection{TP=4}

\begin{table}[H]
    \centering
    \begin{tabular}{|c|c|c|c|c|}
        \hline
        M & K1, N1, N2         & Naive Algorithm (ms) & TP Aware Algorithm (ms) & Speedup \\ \hline
        1 & (8192, 28672, 8192)& 0.472            & 0.282               & 1.67x    \\ \hline
        2 & (8192, 28672, 8192)& 0.512            & 0.286               & 1.79x    \\ \hline
        4 & (8192, 28672, 8192)& 0.513            & 0.287               & 1.79x    \\ \hline
        8 & (8192, 28672, 8192)& 0.518            & 0.285               & 1.82x    \\ \hline
       16 & (8192, 28672, 8192)& 0.512            & 0.286               & 1.79x    \\ \hline
    \end{tabular}
    \captionof{table}{Naive and TP Aware Algorithm Latencies for Various Batch Sizes (Llama-70B, TP=4, A100)}
    \label{tab:naive_vs_tp_various}
\end{table}

\begin{table}[ht]
    \centering
    \begin{tabular}{|c|}
        \hline
        Average Speedup \\ \hline
        1.78x           \\ \hline
    \end{tabular}
    \caption{Average Speedup for Llama-70B, TP=4, A100}
    \label{:tab:avg_speedup_various}
\end{table}

\begin{table}[H]
    \centering
    \begin{tabular}{|c|c|c|c|c|}
        \hline
        M & K1, N1, N2         & Naive Algorithm (ms) & TP Aware Algorithm (ms) & Speedup \\ \hline
        1 & (8192, 28672, 8192)& 0.258            & 0.192               & 1.34x    \\ \hline
        2 & (8192, 28672, 8192)& 0.275            & 0.192               & 1.43x    \\ \hline
        4 & (8192, 28672, 8192)& 0.273            & 0.193               & 1.41x    \\ \hline
        8 & (8192, 28672, 8192)& 0.278            & 0.197               & 1.41x    \\ \hline
       16 & (8192, 28672, 8192)& 0.281            & 0.198               & 1.42x    \\ \hline
    \end{tabular}
    \caption{Naive and TP Aware Algorithm Latencies for Various Batch Sizes (Llama-70B, TP=4, H100)}
    \label{tab:naive_vs_tp_various}
\end{table}

\begin{table}[H]
    \centering
    \begin{tabular}{|c|}
        \hline
        Average Speedup \\ \hline
        1.40x           \\ \hline
    \end{tabular}
    \caption{Average Speedup for Llama-70B, TP=4, H100}
    \label{tab:avg_speedup_various}
\end{table}

\subsection{TP=8}

\begin{table}[H]
    \centering
    \begin{tabular}{|c|c|c|c|c|}
        \hline
        M & K1, N1, N2         & Naive Algorithm (ms) & TP Aware Algorithm (ms) & Speedup \\ \hline
        1 & (8192, 28672, 8192)& 0.495            & 0.284               & 1.74x    \\ \hline
        2 & (8192, 28672, 8192)& 0.503            & 0.276               & 1.82x    \\ \hline
        4 & (8192, 28672, 8192)& 0.539            & 0.291               & 1.85x    \\ \hline
        8 & (8192, 28672, 8192)& 0.530            & 0.286               & 1.85x    \\ \hline
        16 & (8192, 28672, 8192)& 0.512            & 0.286              & 1.79x    \\ \hline
    \end{tabular}
    \captionof{table}{Naive and TP Aware Algorithm Latencies for Various Batch Sizes (Llama-70B, TP=8, A100)}
    \label{tab:naive_vs_tp_various}
\end{table}

\begin{table}[ht]
    \centering
    \begin{tabular}{|c|}
        \hline
        Average Speedup \\ \hline
        1.81x           \\ \hline
    \end{tabular}
    \caption{Average Speedup for Llama-70B, TP=8, A100}
    \label{:tab:avg_speedup_various}
\end{table}

\begin{table}[H]
    \centering
    \begin{tabular}{|c|c|c|c|c|}
        \hline
        M & K1, N1, N2         & Naive Algorithm (ms) & TP Aware Algorithm (ms) & Speedup \\ \hline
        1 & (8192, 28672, 8192)& 0.245            & 0.144               & 1.70x    \\ \hline
        2 & (8192, 28672, 8192)& 0.256            & 0.146               & 1.75x    \\ \hline
        4 & (8192, 28672, 8192)& 0.257            & 0.144               & 1.78x    \\ \hline
        8 & (8192, 28672, 8192)& 0.258            & 0.145               & 1.78x    \\ \hline
        16 & (8192, 28672, 8192)& 0.266           & 0.149              & 1.78x    \\ \hline
    \end{tabular}
    \caption{Naive and TP Aware Algorithm Latencies for Various Batch Sizes (Llama-70B, TP=8, H100)}
    \label{tab:naive_vs_tp_various}
\end{table}

\begin{table}[H]
    \centering
    \begin{tabular}{|c|}
        \hline
        Average Speedup \\ \hline
        1.76x           \\ \hline
    \end{tabular}
    \caption{Average Speedup for Llama-70B, TP=8, H100}
    \label{tab:avg_speedup_various}
\end{table}

\section{Granite-20B}

\begin{figure}[H]
    \centering
    \begin{minipage}{0.48\textwidth}
        \centering
        \textbf{Latency Granite-20B}\par\medskip
        \includegraphics[width=\textwidth]{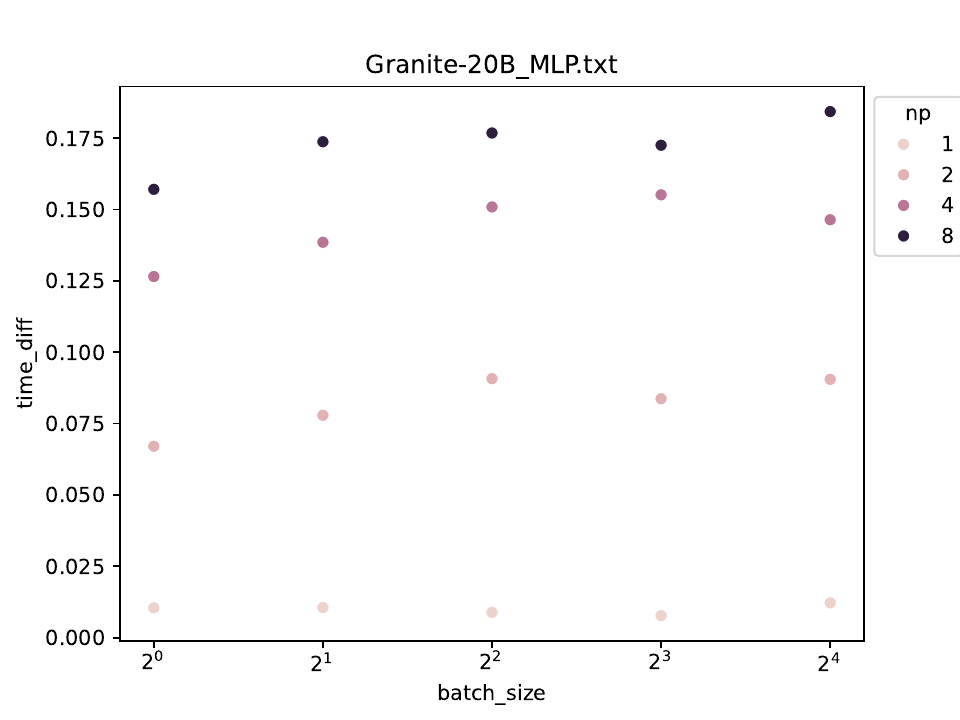}
        \caption{Latency Difference for Granite-20B, A100}
        \label{fig:first_figure}
    \end{minipage}
    \hfill
    \begin{minipage}{0.48\textwidth}
        \centering
        \textbf{Speedup Granite-20B}\par\medskip
        \includegraphics[width=\textwidth]{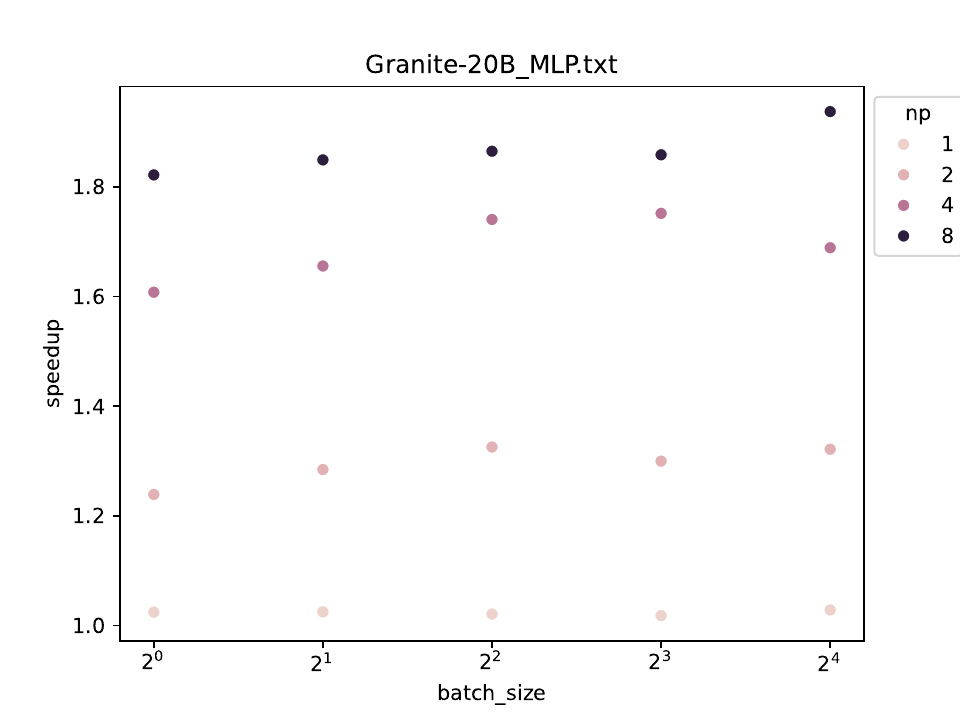}
        \caption{Speedup for Granite-20B, A100}
        \label{fig:second_figure}
    \end{minipage}
\end{figure}

\subsection{Baseline TP=1}

\begin{table}[H]
    \centering
    \begin{tabular}{|c|c|c|c|}
        \hline
        M & K1, N1, N2         & Naive Algorithm (ms) & TP Aware Algorithm (ms) \\ \hline
        1 & (6144, 24576, 6144)& 0.482            & 0.474               \\ \hline
        2 & (6144, 24576, 6144)& 0.476            & 0.471               \\ \hline
        4 & (6144, 24576, 6144)& 0.482            & 0.469               \\ \hline
        8 & (6144, 24576, 6144)& 0.479            & 0.467               \\ \hline
        16 & (6144, 24576, 6144)& 0.487           & 0.475               \\ \hline
    \end{tabular}
    \captionof{table}{Naive and TP Aware Algorithm Latencies for Various Batch Sizes (Granite-20B, TP=1)}
    \label{tab:naive_vs_tp_various}
\end{table}

\begin{table}[H]
    \centering
    \begin{tabular}{|c|c|c|c|}
        \hline
        M & K1, N1, N2         & Naive Algorithm (ms) & TP Aware Algorithm (ms) \\ \hline
        1 & (6144, 24576, 6144)& 0.349            & 0.341               \\ \hline
        2 & (6144, 24576, 6144)& 0.335            & 0.328               \\ \hline
        4 & (6144, 24576, 6144)& 0.325            & 0.319               \\ \hline
        8 & (6144, 24576, 6144)& 0.335            & 0.327               \\ \hline
        16 & (6144, 24576, 6144)& 0.335           & 0.328               \\ \hline
    \end{tabular}
    \caption{Naive and TP Aware Algorithm Latencies for Various Batch Sizes (Granite-20B, TP=1, H100)}
    \label{tab:naive_vs_tp_various}
\end{table}

\subsection{TP=2}

\begin{table}[H]
    \centering
    \begin{tabular}{|c|c|c|c|c|}
        \hline
        M & K1, N1, N2         & Naive Algorithm (ms) & TP Aware Algorithm (ms) & Speedup \\ \hline
        1 & (6144, 24576, 6144)& 0.486            & 0.309               & 1.57x    \\ \hline
        2 & (6144, 24576, 6144)& 0.476            & 0.471               & 1.01x    \\ \hline
        4 & (6144, 24576, 6144)& 0.482            & 0.469               & 1.03x    \\ \hline
        8 & (6144, 24576, 6144)& 0.479            & 0.467               & 1.03x    \\ \hline
        16 & (6144, 24576, 6144)& 0.504            & 0.306              & 1.65x    \\ \hline
    \end{tabular}
    \captionof{table}{Naive and TP Aware Algorithm Latencies for Various Batch Sizes (Granite-20B, TP=2, A100)}
    \label{tab:naive_vs_tp_various}
\end{table}

\begin{table}[ht]
    \centering
    \begin{tabular}{|c|}
        \hline
        Average Speedup \\ \hline
        1.26x           \\ \hline
    \end{tabular}
    \caption{Average Speedup for Granite-20B, TP=2, A100}
    \label{:tab:avg_speedup_various}
\end{table}

\begin{table}[H]
    \centering
    \begin{tabular}{|c|c|c|c|c|}
        \hline
        M & K1, N1, N2          & Naive Algorithm (ms) & TP Aware Algorithm (ms) & Speedup \\ \hline
        1 & (6144, 24576, 6144) & 0.263            & 0.214               & 1.23x    \\ \hline
        2 & (6144, 24576, 6144) & 0.279            & 0.218               & 1.28x    \\ \hline
        4 & (6144, 24576, 6144) & 0.284            & 0.220               & 1.29x    \\ \hline
        8 & (6144, 24576, 6144) & 0.285            & 0.220               & 1.29x    \\ \hline
        16 & (6144, 24576, 6144)& 0.285            & 0.221              & 1.29x    \\ \hline
    \end{tabular}
    \caption{Naive and TP Aware Algorithm Latencies for Various Batch Sizes (Granite-20B, TP=2, H100)}
    \label{tab:naive_vs_tp_various}
\end{table}

\begin{table}[H]
    \centering
    \begin{tabular}{|c|}
        \hline
        Average Speedup \\ \hline
        1.28x           \\ \hline
    \end{tabular}
    \caption{Average Speedup for Granite-20B, TP=2, H100}
    \label{tab:avg_speedup_various}
\end{table}

\subsection{TP=4}

\begin{table}[H]
    \centering
    \begin{tabular}{|c|c|c|c|c|}
        \hline
        M & K1, N1, N2         & Naive Algorithm (ms) & TP Aware Algorithm (ms) & Speedup \\ \hline
        1 & (6144, 24576, 6144)& 0.500            & 0.292               & 1.71x    \\ \hline
        2 & (6144, 24576, 6144)& 0.497            & 0.284               & 1.75x    \\ \hline
        4 & (6144, 24576, 6144)& 0.518            & 0.293               & 1.77x    \\ \hline
        8 & (6144, 24576, 6144)& 0.508            & 0.284               & 1.79x    \\ \hline
       16 & (6144, 24576, 6144)& 0.530            & 0.290               & 1.83x    \\ \hline
    \end{tabular}
    \captionof{table}{Naive and TP Aware Algorithm Latencies for Various Batch Sizes (Granite-20B, TP=4, A100)}
    \label{:tab:naive_vs_tp_various}
\end{table}

\begin{table}[ht]
    \centering
    \begin{tabular}{|c|}
        \hline
        Average Speedup \\ \hline
        1.77x           \\ \hline
    \end{tabular}
    \caption{Average Speedup for Granite-20B, TP=4, A100}
    \label{:tab:avg_speedup_various}
\end{table}

\begin{table}[H]
    \centering
    \begin{tabular}{|c|c|c|c|c|}
        \hline
        M & K1, N1, N2         & Naive Algorithm (ms) & TP Aware Algorithm (ms) & Speedup \\ \hline
        1 & (6144, 24576, 6144)& 0.251            & 0.156               & 1.61x    \\ \hline
        2 & (6144, 24576, 6144)& 0.267            & 0.157               & 1.70x    \\ \hline
        4 & (6144, 24576, 6144)& 0.268            & 0.158               & 1.70x    \\ \hline
        8 & (6144, 24576, 6144)& 0.269            & 0.159               & 1.69x    \\ \hline
        16 & (6144, 24576, 6144)& 0.269           & 0.159              & 1.69x    \\ \hline
    \end{tabular}
    \caption{Naive and TP Aware Model Algorithm for Various Batch Sizes (Granite-20B, TP=4, H100)}
    \label{tab:naive_vs_tp_various}
\end{table}

\begin{table}[H]
    \centering
    \begin{tabular}{|c|}
        \hline
        Average Speedup \\ \hline
        1.68x           \\ \hline
    \end{tabular}
    \caption{Average Speedup for Granite-20B, TP=4, H100}
    \label{tab:avg_speedup_various}
\end{table}

\subsection{TP=8}

\begin{table}[H]
    \centering
    \begin{tabular}{|c|c|c|c|c|}
        \hline
        M & K1, N1, N2         & Naive Algorithm (ms) & TP Aware Algorithm (ms) & Speedup \\ \hline
        1 & (6144, 24576, 6144)& 0.512            & 0.294               & 1.74x    \\ \hline
        2 & (6144, 24576, 6144)& 0.530            & 0.291               & 1.82x    \\ \hline
        4 & (6144, 24576, 6144)& 0.537            & 0.293               & 1.83x    \\ \hline
        8 & (6144, 24576, 6144)& 0.541            & 0.305               & 1.77x    \\ \hline
        16 & (6144, 24576, 6144)& 0.551            & 0.303              & 1.82x    \\ \hline
    \end{tabular}
    \captionof{table}{Naive and TP Aware Algorithm Latencies for Various Batch Sizes (Granite-20B, TP=8, A100)}
    \label{tab:naive_vs_tp_various}
\end{table}

\begin{table}[ht]
    \centering
    \begin{tabular}{|c|}
        \hline
        Average Speedup \\ \hline
        1.80x           \\ \hline
    \end{tabular}
    \caption{Average Speedup for Granite-20B, TP=8, A100}
    \label{:tab:avg_speedup_various}
\end{table}

\begin{table}[H]
    \centering
    \begin{tabular}{|c|c|c|c|c|}
        \hline
        M & K1, N1, N2         & Naive Algorithm (ms) & TP Aware Algorithm (ms) & Speedup \\ \hline
        1 & (6144, 24576, 6144)& 0.252            & 0.148               & 1.70x    \\ \hline
        2 & (6144, 24576, 6144)& 0.255            & 0.142               & 1.79x    \\ \hline
        4 & (6144, 24576, 6144)& 0.259            & 0.141               & 1.84x    \\ \hline
        8 & (6144, 24576, 6144)& 0.257            & 0.140               & 1.84x    \\ \hline
        16 & (6144, 24576, 6144)& 0.255           & 0.140               & 1.82x    \\ \hline
    \end{tabular}
    \caption{Naive and TP Aware Algorithm Latencies for Various Batch Sizes (Granite-20B, TP=8, H100)}
    \label{tab:naive_vs_tp_various}
\end{table}

\begin{table}[H]
    \centering
    \begin{tabular}{|c|}
        \hline
        Average Speedup \\ \hline
        1.78x           \\ \hline
    \end{tabular}
    \caption{Average Speedup for Granite-20B, TP=8, H100}
    \label{tab:avg_speedup_various}
\end{table}

For Llama-70B problem sizes, our TP-Aware method demonstrates an average speedup of 1.22x, 1.78x and 1.81x for TP = 2, 4, 8 respectively on the 8xA100 system. On the 8xH100 system our method demonstrated an average speedup of 1.11x, 1.40x and 1.76x. For Granite-20B problem sizes, we achieve an average speedup of 1.26x, 1.77x and 1.80x on the A100 and 1.28x, 1.68x and 1.78x on the H100.

Notably, as the number of ranks increased so did the corresponding performance improvement. This is the expected result as we expect the naive algorithm communication overhead to increase relatively to overall cost when number of ranks increases. This demonstrates our methods superior scalability over existing methods.

\section{Conclusion}

In this paper we present a novel insight, and discussed it's potential impacts given the increasingly distributed workload of modern day LLM inference. Our contribution is an optimized reordering strategy that minimizes global communication across ranks. We call our method TP-Aware Dequantization. We benchmarked our method on Llama-70B and Granite-20B problem sizes and demonstrated up to 1.81x and 1.76x speedups, respectively, on the NVIDIA DGX A100 and an up to 1.76x and 1.78x speedup on the NVIDIA DGX H100.

\section{Acknowledgements}

We would like to acknowledge important contributions from Less Wright, Meta AI, for providing the H100 experiment results, Hao Yu and Eun Kyung Lee from IBM Research for their review on the presented materiel.

\printbibliography
\end{document}

%% file: equations/reorder_function.tex
\begin{algorithm}
\caption{Reorder Function}
\label{alg:reorder}
\begin{algorithmic}[1]
\Function{Reorder}{$g_{idx\_actorder}$} \Comment{$g_{idx\_actorder}$ is a 1D array containing unordered indices}
    \State $P \gets \Call{ARGSORT}{g_{idx\_actorder}}$ \Comment{$P$ is a 1D array containing the permutation}
    \State  $g_{idx\_optimized} \gets {g_{idx\_actorder}[P]}$ \Comment{Permute $g_{idx\_actorder}$}
    \State \Return $P, g_{idx\_optimized}$ \Comment{$g_{idx\_optimized}$ is a 1D array containing ordered indices}
\EndFunction
\end{algorithmic}
\end{algorithm}

%% file: equations/naive_method.tex

\begin{algorithm}[ht]
\caption{Naive Algorithm}
\label{alg:naive_method_spmd}
\begin{algorithmic}[1]
\Require $X1, W1[P1], W2[P2]$ \Comment{Input activations $X1$ and sharded reordered weight matrices $W1$ and $W2$}
\Require $P1, P2$ \Comment{Permutation arrays}
\Require $rank, size$ \Comment{Processor rank and total number of processors}
\State $Y1_{local} \gets X1_{global}[:, P1] \ @ \ W1_{local}$ \Comment{GEMM}
\State \textcolor{red}{$Y1_{global} \gets \Call{AllGather}{Y1_{local}}$} \Comment{Gather $Y1$ shards from all processors} \label{allgather}
\State $Y1_{global} \gets Y1_{global}[:, P2] $ \Comment{Y1 reordered globally}
\State $Y1_{local} \gets \Call{CHUNK}{Y1_{global}, rank, size, dim=1}$  \Comment{Shard $Y1$ among processors}
\State $Y2_{local} \gets Y1_{local} \ @ \ W2_{local}$ \Comment{GEMM with gathered $Y1$}
\State \textcolor{red}{$Y2_{global} \gets \Call{AllReduce}{Y2_{local}, op=SUM}$} \Comment{Reduce $Y2$ shards across all processors}
\State \Return $Y2_{global}$ \Comment{Each processor has the globally reduced result}
\end{algorithmic}
\end{algorithm}

%% file: equations/optimized_tp_method.tex

\begin{algorithm}[ht]
\caption{TP-Aware Algorithm}
\label{alg:optimized_matmul_spmd}
\begin{algorithmic}[1]
\Require $X1, W1[P1, P2], W2[P2]$  \Comment{Input activations $X1$ and sharded reordered weight matrices $W1$ and $W2$}
\Require $P1$ \Comment{Permutation array}
\Require $rank, size$ \Comment{Processor rank and total number of processors}
\State {$Y1_{local} \gets X1_{global}[:, P1] \ @ \ W1_{local}$} \Comment{GEMM} \label{insight}
\State $Y2_{local} \gets Y1_{local} \ @ \ W2_{local}$ \Comment{GEMM with local $Y1$}
\State \textcolor{red}{$Y2_{global} \gets\Call{AllReduce}{Y2_{local}, op=SUM}$} \Comment{Reduce $Y2$ shards across all processors}
\State \Return $Y2_{global}$ \Comment{Each processor has the globally reduced result}
\end{algorithmic}
\end{algorithm}